\documentstyle[12pt]{article}
\begin{document}
\begin{center}
{\large BOSONIZATION IN THE NONCOMMUTATIVE PLANE }\\
\vskip 2cm
Subir Ghosh\\
\vskip 1cm
Physics and Applied Mathematics Unit,\\
Indian Statistical Institute,\\
203 B. T. Road, Calcutta 700108, \\
India.
\end{center}
\vskip 3cm
{\bf Abstract:}\\
In this Note, we study bosonization of the noncommutative massive Thirring model
in $2+1$- dimensions. We show that, contrary to the duality between massive Thirring model
and Maxwell-Chern-Simons model in ordinary spacetime, in the low energy (or large fermion mass) limit, their noncommutative versions are not
equivalent, in the same approximation.
\vskip 2cm
\noindent
Keywords: $2+1$-Dimensional bosonization, Noncommutative field theory.

\newpage
The string inspired  Non-Commutative (NC) spacetime  \cite{sw} and the subsequent
noncommutative field theories living in the $D$-branes \cite{rev} have remodelled a
number of established ideas of quantum field theories in ordinary spacetime. From a string theory perspective, NC field theories (and NC gauge theories in particular) yield an effective theory for strings in the presence of a large background $B$-field. However, the advantage of the NC field theory formalism is that it deviates very little from the field theory in ordinary spacetime, as far as computational techniques are concerned. Hence, working in NC field theory, some  string theoretic results, albeit in certain limits, are recovered by conventional quantum field theoretic computations. Generically it will be  much harder to obtain analogous results from a string theoretic analysis.

Quite apart from the above mentioned string theory connection, NC quantum field theories are being intensly studied because of some surprising consequences of noncommutativity, such as the ultraviolet-infrared mixing \cite{min}, dipolar behaviour of the excitations in electromagnetic interaction \cite{jab}, a specific type of non-locality \cite{sus}, etc. to name a few. Another interesting observation is that the (noncommutativity parameter) $\theta \rightarrow 0$ limit is not always smooth \cite{sing}, that is results in NC spacetime  do {\it{not}} always reduce to  ordinary spacetime results for $\theta \rightarrow 0$. 
This might seem unexpected since the dynamical variables of NC and ordinary spacetimes are related explicitly through the Seiberg-Witen Map \cite{sw}, through a perturbative expansion in $\theta$. Actually the singularity in $\theta$ appears in the quantum theory when the $\theta \rightarrow 0$ limit and regularization prescription limits do not commute. We will come to the last point later.

A powerful tool in the conventional quantum field theory is the concept of duality (or equivalence) between apparantly dissimilar theories. Apart from the esthetic satisfaction of unifying various theories under one idea, a tangible outcome of duality is that the dual models can represent a physical phenomenon in different limiting domains, such as strong and weak coupling limit etc. It is quite natural to question the fate of a particular duality when the spacetime becomes noncommutative. In the present Note, we will discuss one such duality, {\it {i.e.}} Bosonization or the fermion-boson duality in $2+1$-dimensional NC spacetime.

Bosonization in $1+1$-dimensions dates back to Coleman \cite{col} who showed that that the Sine-Gordon model of a scalar field is dual to the massive Thirring model of self interacting fermions. Subsequently the explicit operator realization of the fermion-boson mapping was provided by Mandelstam \cite{man}. An interesting and useful feature of bosonization is that quantum effects corresponding to the fermionic theory get included in the bosonized effective action, which can be studied classically.

However, generalization of bosonization to higher dimensions is not as complete as in $1+1$-dimensions, (where some simplifications occure due to the topology of the single spatial coordinate). In $2+1$-dimensions, following the ideas in \cite{pol}, Deser and Redlich \cite{red} first studied the equivalence between effective electromagnetic interaction of the $CP^1$ model and a charged massive fermion in powers of inverse fermion mass. Bosonization of the massive Thirring model in the long wavelength regime - the case relevant to us - was considered later by Fradkin and Schaposnik and by Banerjee \cite{2+1}.
The Thirring model, in the lowest non-trivial order in inverse fermion mass, becomes equivalent to the topologically massive $U(1)$ gauge theory, the Maxwell-Chern-Simons theory. The latter model is of interest {\footnote {The vector and tensor gauge theories in $2+1$-dimensions are related to high temperature behaviour of four-dimensional models \cite{wein}.}} in its own right and has been  studied exhaustively in \cite{dj1,dj}. It possesses a single, parity violating, massive, spin one excitation. In the above instances, the parity violation in the bosonic theory, in the form of the Chern-Simons term, comes from the parity violating fermion mass term. Notice that the model enjoys gauge invariance even though the gauge boson is massive, the reason being the topological generation of the mass.

This motivates us to the present work - the study of bosonization of the massive Thirring model, in $2+1$-dimensional NC spacetime. {\footnote {The analogue \cite{nun} of our work in $1+1$-dimensions reveals that the strong-weak coupling duality, similar to the ordinary spacetime result \cite{col,man}, remains intact. The noncommutativity induces a Wess-Zumino-Witten term in the effective $U(1)$ theory.}} In ordinary spacetime, the Massive
Thirring (MT) model - a massive fermionic theory with four fermion
current-current (Thirring) self interaction - is equivalent to the
Self-Dual (SD) model - a massive bosonic theory with Chern-Simons
term as the kinetic term - in the lowest non-trivial order of the
inverse fermion mass \cite{2+1}. The large fermion mass approximation is equivalent to the low energy or long wave length limit of the massive Thirring model, where the fermi-bose transmutation is valid. The Thirring coupling constant gets related to the inverse of the gauge boson mass. Furthermore, due to the
equivalence between the SD and Maxwell-Chern-Simons (MCS) models
\cite{dj,dj1}, the MT model (with no manifest local gauge
invariance)  becomes equivalent to the (manifestly gauge
invariant) MCS model. Our aim is to study,  (a): whether
bosonization of the Non-Commutative MT (NCMT) theory along the lines of ordinary
spacetime is possible and (b): if it is so, whether the NCMT-NCMCS
duality is preserved.

Our results consist of a good
news and a bad news. The good news is that, (in analogy with ordinary spacetime \cite{2+1}),
{\it {the NCMT model can be bosonized in powers of the inverse fermion mass}}.
The bad news is that, (contrary to the ordinary spacetime \cite{2+1}),
{\it {the duality
between the NCMT  and NCMCS
models is lost, even in the large fermion mass limit}}. The reason is the following.
Recently it was shown by us \cite{sg} that the NCSD-NCMCS duality survives.
However, here we show that bosonization of the NCMT theory induces a theory which differs from
the NCSD theory studied previously \cite{sg}. Hence the NCMT-NCMCS duality is lost. These
constitute the main results of this paper, schematically summarized below:
$$Ordinary~spacetime~~~~~~MT~(fermion)\approx SD~(boson) \approx MCS~(boson)$$
$$ \Rightarrow MT~(fermion) \approx MCS~(boson) ,$$
$$NC~spacetime~~~NCMT~(fermion)\approx Bosonic ~ model \neq NCSD~(boson) \approx NCMCS~(boson) $$
$$\Rightarrow NCMT~(fermion) \neq NCMCS~(boson) .$$

After putting our work in its proper perspective, we now move on to explicit computations. The spacetime is characterized by a noncommutativity of the form,
\begin{equation}
[x^{\rho},x^{\sigma}]_{*}=i\theta^{\rho\sigma},
\label{nc}
\end{equation}
where the ordinary product is replaced by the  Moyal-Weyl or $*$ product,
\begin{equation}
\hat p(x)*\hat q(x)=e^{\frac{i}{2}\theta_{\mu\nu}\partial_{\sigma_{\mu}}\partial_{\xi_{\nu}}}\hat p(x+\sigma )\hat q(x+\xi )\mid_{\sigma =\xi =0}
=\hat p(x)\hat q(x)+\frac{i}{2}\theta^{\rho\sigma}\partial_{\rho}\hat p(x)\partial_{\sigma}\hat q(x)+~O(\theta^{2}).
\label{mw}
\end{equation}
The {\it hatted} variables live in NC spacetime. Generally $\theta^{\rho\sigma}$ is taken to be
a constant tensor, but this need not always be the case \cite{ssg}.
In this paper we will focus on $2+1$-dimensional NC spacetime.

Let us discuss NC bosonization first. The NCMT model is,
$$
\hat S_{Th}=\int d^{3}x ~[\bar {\hat\psi}(x)*(i\gamma^{\mu}\partial_{\mu}+m)\hat \psi(x)-\frac{g^{2}}{2}\hat j^{\mu}(x)*\hat j_{\mu}(x)]$$
\begin{equation}
=\int d^{3}x ~[\bar {\hat\psi}(x)(i\gamma^{\mu}\partial_{\mu}+m)\hat \psi(x)-\frac{g^{2}}{2}\hat j^{\mu}(x)\hat j_{\mu}(x)],
\label{thr}
\end{equation}
where the fermion current $\hat j_{\mu}$is defined as
\begin{equation}
\hat j^{\mu}(x)=\bar {\hat\psi}(x)*\gamma^{\mu}\hat\psi (x).
\label{cur}
\end{equation}
The second equality in (\ref{thr}) follows from the property of $*$-product under integral.
The next step is to compute the effective action by integrating out the fermions.
We consider an alternative action,
$$
\hat S=\int d^{3}x ~[\bar {\hat\psi}(x)*(i\gamma^{\mu}\partial_{\mu}+m)\hat \psi(x)+\hat j^{\mu}(x)*\hat B_{\mu}(x)+\frac{1}{2g^{2}}\hat B^{\mu}*\hat B_{\mu}]$$
\begin{equation}
=\int d^{3}x ~[\bar {\hat\psi}(x)(i\gamma^{\mu}\partial_{\mu}+m)\hat \psi(x)+\hat j^{\mu}(x)\hat B_{\mu}(x)+\frac{1}{2g^{2}}\hat B^{\mu}\hat B_{\mu}],
\label{aux}
\end{equation}
where the Thirring interaction is linearized by introducing a field $\hat B_{\mu}$.
This is similar to the formalism followed in ordinary spacetime \cite{2+1}.
However,
there is a subtlety involved, which is peculiar to NC spacetime. Depending on the positioning
of $\hat B_{\mu}$ and $\hat \psi$, the covariant derivative can act in three ways \cite{gs},
\begin{equation}
\hat D_{\mu}\hat\psi 
\begin{array}[t]{l}
 =\partial_{\mu}\hat\psi+i\hat B_\mu*\hat\psi \\
=\partial_{\mu}\hat\psi-i\hat\psi*\hat B_\mu \\
=\partial_{\mu}\hat\psi+i(\hat\psi*\hat B_\mu -\hat B_\mu*\hat\psi )
\end{array}
\label{cov}
\end{equation}
which are termed respectively as fundamental, anti-fundamental and adjoint representations.
Notice that we have chosen the anti-fundamental one in (\ref{aux}), since this will reduce
to the original
Thirring model (\ref{thr}) once $B_{\mu}$ is integrated out. 

We now follow the work of Grandi and Silva \cite{gs} in computing the fermion determinant and
write down the effective action,
\begin{equation}
\hat S[\hat B]=\int d^{3}x [-\frac{1}{8\pi}\epsilon^{\mu\nu\lambda}(\hat B_{\mu}*\partial_{\nu}\hat B_{\lambda}+\frac{2i}{3}\hat B_{\mu}*\hat B_{\nu}*\hat B_{\lambda})+\frac{1}{2g^{2}}\hat B^{\mu}\hat B_{\mu}] +O(\frac{1}{m}).
\label{bosz}
\end{equation}
Pauli-Villars regularization has been invoked and only the parity violating contribution is
exhibited. The first term is the NC Chern-Simons term. The details of the derivation are to be found in \cite{gs}. This completes the first part of our
work, that is bosonization of NCMT {\footnote{Referring to our earlier comment on the smoothness of the $\theta \rightarrow 0$ limit, notice that even though the coupling in the adjoint representation vanishes for $\theta =0$, the effective action is non-zero \cite{gs}. This is relevant for Majorana fermions which are neutral in ordinary spacetime. However, this does not concern the present work.}}.

To understand the non-existance of NCMT-NCMCS duality, we briefly recall earlier works.
The (ordinary spacetime) duality between the SD model, otained by bosonizing the MT model,
\begin{equation}
S_{SD}=\int d^{3}x[\frac{1}{2g^2}B^\mu B_\mu-\frac{1}{8\pi}\epsilon ^{\alpha\beta\gamma}B_\alpha\partial _\beta B_\gamma ]
\label{sd}
\end{equation}
and the MCS model,
\begin{equation}
 S_{MCS}=\int d^{3}x[-\frac{1}{2}(\partial_{\alpha}A_{\beta}-\partial_{\beta}A_{\alpha})\partial^{\alpha}A^{\beta}+\frac{2\pi}{g^2}\epsilon^{\alpha\beta\gamma}A_{\alpha}\partial_{\beta}A_{\gamma}].
\label{mcs}
\end{equation}
discovered by Deser and Jackiw \cite{dj,dj1},
was surprising since the latter is a gauge theory whereas the former is (naively) not.
However constraints of the theories induce identical spectra and a mapping between degrees
of freedom of the two theories \cite{dj1}. This equivalence was further corroborated
in \cite{dj} where a "`Master"' Lagrangian was constructed, which was capable of generating both
the SD and MCS models.

The duality between the following NC versions of SD and MCS
theories, shown in \cite{sg} by exploiting the "`Master"' Lagrangian technique,
\begin{equation}
\hat S_{SD}=\int d^{3}x[\frac{1}{2g^2}\hat B^\mu *\hat B_\mu-\frac{1}{8\pi}\epsilon ^{\alpha\beta\gamma}\hat B_\alpha *\partial _\beta \hat B_\gamma ]
\label{ncsd}
\end{equation}
\begin{equation}
\hat S_{MCS}=\int d^{3}x[-\frac{1}{2}(\partial_{\alpha}(\hat A +\hat a )_{\beta}-\partial_{\beta}(\hat A +\hat a )_{\alpha})*\partial^{\alpha}(\hat A +\hat a )^{\beta}+\frac{2\pi}{g^2}\epsilon^{\alpha\beta\gamma}(\hat A +\hat a )_{\alpha}*\partial_{\beta}(\hat A +\hat a )_{\gamma}].
\label{ncmcs}
\end{equation}
is all the more non-trivial since (\ref{ncmcs}) is generated via the (inverse)
Seiberg-Witten map {\footnote {The significance of the Seiberg-Witten map is that under an NC or $*$-gauge
transformation of $\hat A_{\mu}$ by,
$$\hat\delta \hat A_{\mu}=\partial_{\mu}\hat \lambda +i[\hat\lambda ,\hat A_{\mu}]_{*},$$
$A_{\mu}$ will undergo the transformation $$\delta A_\mu
=\partial_{\mu}\lambda . $$ Subsequently, under
this mapping, a gauge invariant object in conventional spacetime
will be mapped to its NC counterpart, which will be $*$-gauge
invariant.}} \cite{sw}, which, valid to the first non-trivial order in $\theta$, is
$$
A_{\mu}=\hat A_{\mu}-\theta^{\sigma\rho}\hat A_{\rho}(\partial_{\sigma}\hat A_{\mu}-\frac{1}{2}\partial_{\mu}\hat A_{\sigma})\equiv \hat A_{\mu}+\hat a_{\mu}(\hat A_{\nu},\theta )
$$
\begin{equation}
\lambda =\hat \lambda +\frac{1}{2}\theta^{\rho\sigma}\hat A_{\rho}\partial_{\sigma}\hat \lambda .
\label{swm}
\end{equation}
As stated before, the "`hatted"' variables on the right are NC degrees of freedom and gauge transformation parameter. It should be mentioned that the $O(\theta )$ expression of the Seiberg-Witten map is used only because the higher order terms in $\theta$ can not be obtained uniquely \cite{fid}. However, it is important to note that the equivalence result remains valid to all orders in $\theta$ as the explicit form of the map is not required in this proof. This is discussed in \cite{tin}. Indeed, the $O(\theta)$ analysis plays a central role since in NC spacetime physics as most of the results till date refer to $O(\theta)$ corrections over the results in normal spacetime.

In fact,
(\ref{ncmcs}) is nothing but the sum of NC Maxwell term and NC Chern-Simons term, correct up
to the first non-trivial order in $\theta$. $S_{MCS}$ in (\ref{mcs}) is transformed to
$\hat S_{MCS}$ in  (\ref{ncmcs}) by using the Seiberg-Witten map \cite{sw} given in (\ref{swm}).
On the other hand, $\hat S_{SD}$ (\ref{ncsd}) is gotten from  $S_{SD}$ (\ref{sd}) simply by replacing the ordinary
products by $*$-product and using $B_{\mu}\equiv \hat B_{\mu}$. It does not require the
Seiberg-Witen map. Notice that for this reason, the parity-odd term in (\ref{ncsd}) is not
the NC extension of the Chern-Simons term. The non-abelian extension of these ideas are discussed in \cite{nab1} in ordinary spacetime and
in \cite{nab2} in NC spacetime.

The difference between the mechanisms by which $S_{SD}\rightarrow \hat S_{SD}$
$((\ref{sd})\rightarrow (\ref{ncsd}))$ and $S_{MCS}\rightarrow \hat S_{MCS}$
$((\ref{mcs})\rightarrow (\ref{ncmcs}))$ are obtained, is due to the fact that since
$S_{MCS}$ has a gauge invariance, $\hat S_{MCS}$ must have the corresponding $*$-gauge
invariance. This is ensured by invoking the Seiberg-Witten map. On the other hand, $S_{SD}$ is not a
manifestly gauge invariant theory the Seiberg-Witten map does not come into play in the generation of its NC
version.

It is now straightforward to see that even in the lowest
non-trivial order in $\theta$, the NCSD model given in
(\ref{ncsd}) and the theory (\ref{bosz}) obtained from
bosonization of the NCMT model are different, because of the triple $B_{\mu}$ term in (\ref{bosz}). Note that this difference vanishes in ordinary spacetime. This proves that the NCMCS theory is
not the dual of NCMT model. This constitutes the
second part of our statement advertised before.

A comment about nomenclature is possibly in order. This pertains
to the fact that which of the models between (\ref{ncsd}) and
(\ref{bosz}) should be referred to as the NCSD. One can argue in
favor of (\ref{bosz}) since starting from the ordinary spacetime
SD model, in (\ref{bosz}), the Chern-Simons term is
generalized to its NC version and the mass term remains as such.
On the other hand we contend \cite{sg} that (\ref{ncsd}) should be
termed as the NCSD model since the SD model being gauge variant,
its NC version should be obtainable by only converting the
ordinary products to $*$-products. Also, (\ref{ncsd}) obeys the
self dual equation whereas (\ref{bosz}) does not. All the same, keeping this ambiguity
aside, the fact remains that the model (\ref{bosz}) obtained by bosonizing the NCMT is
different from the model (\ref{ncsd}) and in \cite{sg} the latter was shown to be equivalent
to the NCMCS model.

To conclude, we have shown that, (keeping in mind the subtleties
involved in noncommutative field theories), the noncommutative
massive Thirring model can be bosonized in the large fermion mass
limit, along the lines of its ordinary spacetime version. However,
unlike the ordinary spacetime result, the duality between
noncommutative massive Thirring model and noncommutative
Maxwell-Chern-Simons model does not survive. Intuitivly, the reason for this failure is also not hard to guess. The noncommutative  gauge theories that appear here are structurally akin to non-abelian gauge theories in ordinary spacetime and from previous experience \cite {nab1} we know that these dualities are to be understood in a more restricted sense in the non-abelian setup.

\vskip 1cm 
\noindent
{{\bf Acknowledgement:}} It is a
pleasure to thank Dr. Rabin Banerjee for correspondence.

\end{document}